\documentclass[12pt]{iopart}
\pdfoutput=1
\usepackage{color}
\usepackage{iopams}
\usepackage{graphicx}
\usepackage{subfigure}

\usepackage{lineno}

\newcommand{\ds}{\displaystyle}
\newcommand{\oh}{\frac{1}{2}}

\begin{document}
 \title{ The effect of quenched disorder in neutral theories}
 \author{Claudio Borile}
 \address{Dipartimento di Fisica `G. Galilei', Universit\`a di Padova, Via Marzolo 8, 35131 Padova, Italy}
 \author{Amos Maritan}
 \address{Dipartimento di Fisica `G. Galilei', Universit\`a di Padova, CNISM \& INFN, Via Marzolo 8, 35131 Padova, Italy}
 \author{Miguel A. Mu\~noz}
 \address{ Departamento de
    Electromagnetismo y F{\'\i}sica de la Materia e Instituto Carlos I
    de F{\'\i}sica Te\'orica y Computacional. Universidad de Granada.
    E-18071, Granada, Spain}

 \begin{abstract}
   We study systems with two symmetric absorbing states, such as the voter model and variations of it, which have been broadly used as minimal neutral models in genetics, population ecology, sociology, etc. We analyze the effects of
   a key ingredient ineluctably present in most real applications: random-field-like quenched disorder. In accord with simulations and previous findings, coexistence between the two competing states/opinions turns out to be strongly favored by disorder in the standard voter model; actually, a disorder-induced phase transition is generated for any finite system size in the presence of an arbitrary small spontaneous-inversion rate (preventing absorbing states from being stable).  For non-linear versions of the voter model a general theory (by AlHammal et al.) explains that the spontaneous breaking of the up/down symmetry and an absorbing state phase transition can occur either together or separately, giving raise to two different scenarios.  Here, we show that he presence of quenched disorder in non-linear voter models does not allow the separation of the up-down (Ising-like) symmetry breaking from the active-to-absorbing phase transition in low-dimensional systems: both phenomena can occur only simultaneously, as a consequence of the well-known Imry-Ma argument generalized to these non-equilibrium problems. When the two phenomena occur at unison, resulting into a genuinely non-equilibrium (``Generalized Voter'') transition, the Imry-Ma argument is violated and the symmetry can be spontaneously broken even in low dimensions.
 \end{abstract}
 \maketitle
\section{Introduction}

In Nature, thermodynamic equilibrium is the exception rather than the rule. Almost the entire world of observable collective phenomena --from the formation of galaxies and stars to the self organization of communities of insects or neurons in the brain-- can be best adressed within an out-of-equilibrium framework \cite{Zwanzig}. Owing to this, huge efforts have been devoted to the study of the statistical mechanics of nonequilibrium systems and their applications in physics, biology, ecology etc. Understanding the fundamental laws governing the statistics and the dynamics of these systems is a formidably complex task; still, a great variety of them --even if \emph{a priori} very different in their nature-- show similar underlying statistical features that can be captured by rather simple probabilistic toy models, a few of which have become paradigmatic \cite{LiggettInteracting,Marro,Henkel,Odor}.

The Voter Model (VM) \cite{LiggettInteracting} occupies a preponderant position in this context. It was independently proposed in various research fields to study neutral genetic drift in an ideal population \cite{Kimura69, Crow}, the competition for territory between two countries \cite{Clifford73}, spreading of infectious diseases \cite{Pinto} or language competition. It is one of the few known interacting-particle models which is exactly solvable in any spatial dimension \cite{LiggettInteracting,LiggettMarkov} and is now extensively employed and studied --even if with different names-- in population genetics \cite{Blythe07}, ecology \cite{Hubbell, Durrett94}, sociology \cite{Castellano09b}, and linguistics \cite{Croft02, Croft06} among other disciplines.  The VM is simply defined: In a community of individuals (voters) arranged on the vertices of a regular lattice (or, more generally, upon an arbitrary network \cite{Castellano09b}) each voter has one out of two equivalent states (opinions) --generalizations to more states being straightforward-- and at every discrete time step the dynamics is defined as follows:
 \begin{enumerate}
 \item  a voter is randomly selected,
 \item its opinion is substituted by the one of a randomly chosen nearest neighbor,
 \item the process is iterated \emph{ad infinitum} or until consensus (all voters sharing the same opinion) is reached.
\end{enumerate}
Some relevant features  which  determine the global behavior of the model and its observables are \cite{LiggettInteracting,Blythe07,Dornic01,AlHammal05}:
\begin{itemize}
\item the dynamics is completely \emph{neutral}, that is, the opinions' labels can be switched leaving the dynamics unaltered, this makes the VM the most basic model to analyze neutral theories in population genetics (where the states correspond to ``alleles'') and in ecology (in which the states are different ``species'');

\item it is a linear model in the sense that the probability for a single voter to change its opinion, that is, the \emph{flipping probability} $f$, increases \emph{linearly} with the number of discordant neighbor voters. This allows for an exact moment-closure of the dynamical equations for the $n-$point correlators;

\item it has no free parameters and since it lacks of any type of characteristic (length or time) scale it seats, by definition, at a critical point;

\item it is characterized by a purely noise-driven diffusive dynamics without surface tension at the boundaries between different domains of opinion;

\item  it exhibits two $\mathbb{Z}_2$-symmetric absorbing states, from which the system cannot escape.
\end{itemize}
Some key properties can be summarized as follows \cite{Krapivsky}: For lattices of dimension $D\leq 2$ the system reaches one of its two absorbing states with probability one in the thermodynamic limit; instead, for $D>2$ the active state --with two coexisting opinions -- lasts indefinitely.  For finite-size systems consensus is eventually reached almost surely, i.e. with probability one, since it is the only state in which the dynamics ceases. If $N$ is the total number of voters, the mean time to reach consensus, $T_N$, scales as $N^2$ in $D=1$, $N\log N$ in $D=2$, and as $N$ for $D\geq 3$. Similarly, in the infinite-size limit, the 2-point correlation function $G(r, t)$, measuring the probability that two voter at distance $r$ at time $t$ have the same opinion, approaches $1$ at arbitrary fixed distance $r$ at large times for $D\leq2$, while it goes $G(r, t) \sim r^{2-D}$ for $D>2$, consistent with the lack of consensus.

Nonlinear generalizations of the VM have been considered, for example, in neutral ecology, where the VM is the simplest model mimicking stochastic species competition.  Nonlinearity in this case stems from a dependence of the dynamical rules on the density of individuals of each specific species in a local neighborhood \cite{Molofsky99,Neuhauser99,Chesson00,Schweitzer09}. Actually, it is well understood that negative density-dependence (implying that a locally infrequent species has a competitive advantage) significantly favors species coexistence. Similarly, a set of models labeled ``with heterozigosity selection'' where introduced in population genetics (having locally different coexisting ``alleles'' is favored or, in other words, minorities tend to be preserved) \cite{Sturm08}.  Different non-linear variations of the VM have been reported to exhibit a novel type of phase transition, lying in the so-called \emph{Generalized Voter} (GV) universality class (see \cite{Dornic01, AlHammal05,Castellano09a,Lee10,Canet} and refs. therein).  This class is characterized by a critical point separating and active from an absorbing region, whose scaling right at criticality coincides with that of the original VM.  Moreover, its dynamics is dual to a family of models of branching and annihilating random walkers with parity number conservation \cite{LiggettMarkov, Peliti86}.

A key ingredient likely to strongly influence the dynamics of real systems is \emph{quenched disorder}.  Disorder --which is unavoidable in Nature-- can be defined na\"{\i}vely as an intrinsic component of randomness in the interaction among the ``micro-constituents'' or in the topological structure over which the dynamics takes place. It is well known from statistical mechanics that quenched disorder can have a dramatic impact in equilibrium systems \cite{Harris74,Grinstein76,Imry79,Berker93}.  Recently, there has been a growing interest in studying the effects of disorder in genuinely nonequilibrium models and in particular in models with absorbing states \cite{Moreira96,Cafiero98,Hoo03,Odor06}.  Frachebour, Krapivsky, and Redner \cite{Krapivsky95} studied the influence of quenched disorder in the form of impurities for a model of catalysis with two symmetric absorbing states, showing that a non-trivial steady state emerge.  More recently, Masuda et al. \cite{Masuda10, Masuda11} showed that quenched (random-field like) disorder --creating an intrinsic preference of each individual for a particular state/opinion-- hinders the formation of consensus, hence favoring coexistence.  Actually, the presence of just a few different ``zealots'' --not allowed to change their intrinsic state-- suffices to prevent consensus \cite{Mobilia07}. Along similar lines, Pigolotti and Cencini \cite{Pigolotti10} analyzed in the context of neutral ecology a version of the VM in which at each location there is an intrinsic preference for one particular species, leading to mixed states (no consensus/monodominance) lasting for times that grow exponentially with system size.  By studying a similar model, Barghathi and Vojta \cite{Vojta12} have very recently stressed that --contrarily to what happens in equilibrium systems, where a well known (Imry-Ma) argument precludes symmetries to be spontaneously broken in low-dimensional systems in the presence of quenched random fields \cite{Imry75}, intrinsically non-equilibrium phase transitions, such as those in the GV class, do persist in low-dimensional systems ($D=1$) in the presence of random fields, even if with a different type of critical behavior \cite{Vojta12}.
Despite of all these results, a complete and coherent theoretical framework to understand the effects of disorder in VM-like systems is still missing.  

In this paper, aimed at shedding further light on this problem, we consider a VM where each voter experiences an intrinsic tendency to align with a particular opinion. We propose a systematic study of this type of models with an approximate analytical description of the underlying stochastic dynamics complemented with computer simulations. 

The paper is organized as follows: In the next section we introduce a model with two symmetric absorbing states and quenched random fields. In section 3 we first discuss the mean-field (deterministic) limit of the model and obtain an approximate solution taking into account finite-size (stochastic) effects. We find that disorder keeps the system away from the absorbing boundaries and --in agreement with previous findings-- the typical time to fixation is exponential in the system size instead of linear as in the VM.
Section 4 discusses the connections with the GV universality class. Section 5 extends the analysis to a class of nonlinear voter model, while in Section 6 concluding remarks and considerations are presented.

\begin{figure}[htp]
 \centering
 \includegraphics[width=0.85\textwidth]{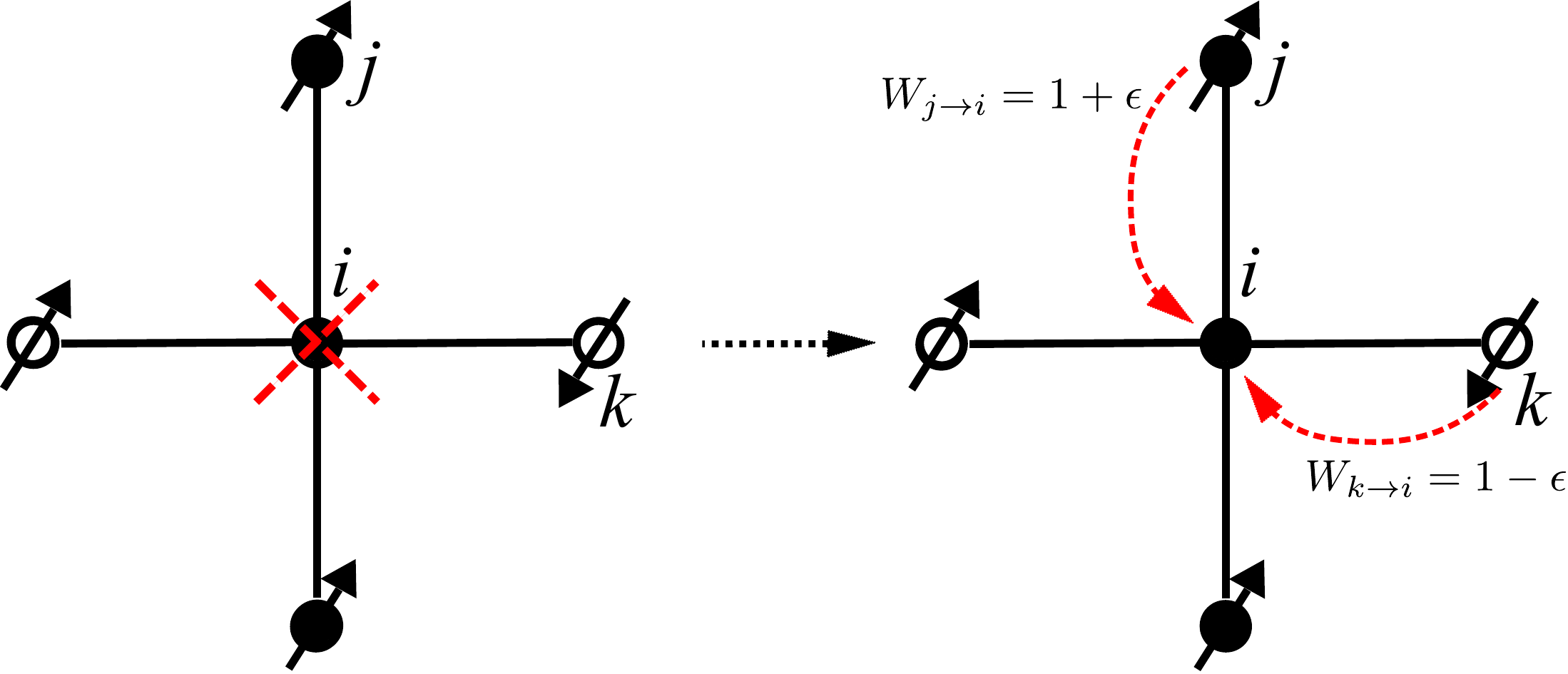}
 \caption{Cartoon of the microscopic dynamical rules of the model. (Left) The underlying lattice is represented by black/white nodes representing the quenched random field pointing upward/downward. A central ``spin'', $i$, with associated random field pointing upward is removed and replaced by (Right) a nearest-neighboor state; the relative transition rates to chose any (only two of them are depicted with red dashed arrows) depends on whether it is aligned or not with $\tau_i$: $1 +\epsilon$ if it becomes aligned, $1- \epsilon$ if it does not. }
 \label{fig:cartoon}
\end{figure}

\section{The model} 
\subsection{Definition of the model and notations} 
We consider a Voter Model defined on a $D$-dimensional lattice ($\Lambda\subset \mathbb{Z}^D$) with $N=L^D$ sites, denoted by $i,j,\dots$ \cite{LiggettInteracting, LiggettMarkov}. At each site $i$ resides a binary or spin variable $\sigma_i\in \{+1,-1\}$ and a random binary field $\tau_i$ to which $\sigma_i$ is locally coupled, favoring its alignment with the field.  The values of $\tau$ are quenched, that is, a particular realization of the disorder is extracted and does not change during the dynamics.  To ensure the global up-down (plus/minus) symmetry we partition the lattice into two disjoint sets of the same size, namely $\Lambda=\Lambda^+\sqcup\Lambda^-$, with $\Lambda^\pm=\{i\in \Lambda\ :\ \tau_i=\pm 1\}$ and $|\Lambda^+|=|\Lambda^-|$.  In finite spatial dimensions, for large enough systems, this constraint can be relaxed to be satisfied just on average over the lattice: we take $\tau_i$ as \emph{i.i.d.} random variables taking values in $\{+1, -1\}$ with uniform probability. At each site the coupling-strength between the spin and the random field is controlled by the free parameter $\epsilon\in [0,1]$, where $\epsilon=0$ stands for the uncoupled (pure VM) case and $\epsilon=1$ implies that each spin remains frozen in the direction of its random field.  The model is completely defined --in its continuous-time version-- by specifying the transition rates $W$ for a generic spin $i$, see Figure \ref{fig:cartoon}, namely: 
\begin{equation} 
W(\sigma_i\to-\sigma_i) = \displaystyle \frac{1-\epsilon\tau_i\sigma_i}{2z}\sum_{j\in \partial i}(1-\sigma_i\sigma_j), 
\label{eq:model} 
\end{equation} 
where $\partial i$ is the set of nearest neighbors of $i$ and $z$ is the lattice coordination number ($z=2D$, for a regular square lattice). The term independent of $\epsilon$ describes the standard VM dynamics \cite{LiggettMarkov}, while the second term is proportional to $\epsilon$ and the flipping probability is enhanced or reduced depending on whether the spin becomes aligned or not with its random field.

\subsection{Mapping onto a birth-death Fokker-Planck equation} 
As a first step to construct a mean-field solution, let us consider the dynamics on a complete graph, that is, each spin connected to every other spin (in eq. (\ref{eq:model}) this corresponds to $\partial i = \{j\neq i\}$). The macroscopic state of the system is univocally determined by the value of two variables, $x$ and $y$, which represent the fraction of up and down spins aligned with their corresponding random fields, respectively:
\begin{equation} \left\{ \begin{array}{l}
 x\equiv \frac{1}{N} \displaystyle \sum_{i=1}^N \mathbb{I}(\sigma_i=\tau_i=+1)\\
 y\equiv \frac{1}{N} \displaystyle \sum_{i=1}^N \mathbb{I}(\sigma_i=\tau_i=-1).
\end{array}
\right. 
\end{equation}
These two variables are defined in the interval $[0,1/2]$ and, since the total number $N$ of spins is constant, the total fraction of up and down spins ($X$ and $Y$ respectively) are readily obtained as $X=1-Y=1/2+x-y\in [0,1]$. The global magnetization is given by $\phi=2(x-y)$.  We can map this spin model onto a birth-death process considering the Master Equation (ME) for $P(x, y, t)$, the joint probability of having at time $t$ a fraction $x$ and $y$ of up and down spins aligned with their local field, respectively.  In this (fully connected or mean-field) version of the model, $P(x, y, t)$ evolves through discrete steps $x\to x'=x\pm 1/N$ and $y\to y'=y\pm 1/N$, with transition rates given by (see \cite{Masuda10,Masuda11,Pigolotti10})
\begin{equation}
 \begin{array}{l}
 W^b_x=W(x\to x+1/N)=(1+\epsilon)(\oh-x)(\oh+x-y)\\
 W^d_x=W(x\to x-1/N)=(1-\epsilon)x(\oh-x+y)\\
 W^b_y=W(y\to y+1/N)=(1+\epsilon)(\oh-y)(\oh+y-x)\\
 W^d_y=W(y\to y-1/N)=(1-\epsilon)y(\oh-y+x),
\end{array}
\label{eq:rates}
\end{equation}
The standard Kramers-Moyal expansion \cite{Gardiner} leads to the Fokker-Planck approximation of the original ME for the evolution in time of $P(x, y, t)$, that we write as
\begin{equation}
 \partial_t P(x, y, t) = \partial_x \left[-A_x P+ \frac{1}{2N}\partial_x(B_xP)\right] +\partial_y \left[-A_y P+ \frac{1}{2N}\partial_y(B_yP)\right],
 \label{eq:FPxy}
\end{equation}
where $A_i=W^d_i-W^b_i$ represent the drift terms and $B_i=W^d_i+W^b_i$ the diffusion terms, $i=x,y$, and time has been rescaled in units of $1/N$.

\section{Steady state Analysis} 

\subsection{Deterministic  limit} 
For the time being we focus only on the limit $N\to \infty$, when the diffusion terms can be safely set to zero and the dynamics becomes deterministic, namely $\dot{x}=A_x$ and $\dot{y}=A_y$. We perform a change of variables \cite{Masuda10, Masuda11} that will be useful for the later analysis by defining $\Sigma\equiv x+y\in [0, 1]$ and $\Delta\equiv x-y \in [-1/2, 1/2]$, from which the global magnetization can be written as $\phi=2\Delta$. In this notation, the deterministic equations become \begin{equation}
 \left\{
 \begin{array}{l}
  \dot{\Delta}=\epsilon\Delta(1-2\Sigma)\\[12pt]
  \dot{\Sigma}=\frac{1}{2}(1+\epsilon)-\Sigma-2\epsilon\Delta^2.
 \end{array}
\right. 
\label{eq:SigmaDelta}
\end{equation}
The analysis of the dynamical system described in Eq. (\ref{eq:SigmaDelta}) gives already some interesting results (as already outlined in \cite{Masuda10,Masuda11,Pigolotti10}). For $\epsilon=0$ one obtains a line of stable fixed points at $\Sigma=1/2$ (and arbitrary $\Delta$), recovering the VM results and hence the system always reaches an absorbing state when fluctuations are considered, also in the infinite-size system. Instead, for $\epsilon>0$, the phase portrait changes dramatically and the line of fixed points breaks into three fixed points: two of them are unstable corresponding to the absorbing states of the VM dynamics, at $\Sigma=1/2$ and $\Delta=\pm1/2$, while the third, at $\Sigma=1/2(1+\epsilon)$ and $\Delta=0$, is stable and corresponds to an active state with zero magnetization, i.e. a phase of symmetric coexistence of the two opinions. Thus the infinite-size limit for this latter case is always in the active phase, but what are the effects of fluctuations when the size is finite? We will answer this question in the next sections.

\subsection{Finite-$N$ approximate solution and role of the stochastic noise}

For finite-size systems fluctuations cannot be neglected and the system is expected to fluctuate around the deterministic stable fixed point. We expect a priori that only a large collective deviation can bring the system to one of the absorbing states which is, nevertheless, ineluctably reached. To study this we consider the Fokker-Planck eq. (\ref{eq:FPxy}). It is easy to verify --by computing cross-derivatives-- that this equation does not admit a potential solution for the stationary probability distribution \cite{Gardiner}. The lack of a stationary potential reflects the intrinsically non-equilibrium nature of the problem. A possible strategy would be to construct non-differentiable non-equilibrium potentials following the strategy in \cite{Graham85, Graham86}. Instead, here, we follow a simpler solution by seeking for a suitable (adiabatic) approximation allowing us to reduce the problem to a one-variable one \cite{Ohtsuki06}.

Let us consider $\epsilon \ll 1$, then Eq. (\ref{eq:SigmaDelta}) have two different characteristic relaxation times: $\Sigma$ relaxes in a time $\mathcal{O}(1)$ whereas $\Delta$ in a much longer time-scale $\mathcal{O}(\epsilon^{-1})$.  Thus, one can assume that the system first relaxes to the nullcline orbit $\dot{\Sigma}=0$ and then the dynamics is constrained to take place uniquely on such a one-dimensional manifold. Consequently, we adopt a quasi-steady-state scheme where the variable $\Sigma$ is substituted by its value in the nullcline orbit, namely 
\begin{equation}
 \Sigma\to \bar{\Sigma}=\oh(1+\epsilon)-2\epsilon\Delta^2
 \label{eq:sigmabar}
\end{equation}
and it is treated like a deterministic quantity, that is, fluctuations in its direction are discarded.

Within this approximation the Fokker-Planck equation for the probability distribution of $\Delta$, or equivalently $\phi$, $\mathcal{P}(\phi, t)$, is obtained with a change of variables in Eq.(\ref{eq:FPxy}) from $(x,y)$ to $(\Delta, \Sigma)$ and substituting the variable $\Sigma$ in the remaining equation with $\bar{\Sigma}$ of Eq. (\ref{eq:sigmabar}). The diffusion term for the variable $\Sigma$ is neglected and we are left with the following $1$-dimensional FP equation \begin{equation}
 \dot{\mathcal{P}}(\phi, t)=-\partial_\phi \left[ \mathcal{A}(\phi)\mathcal{P}(\phi,t)\right]+\frac{1}{2} \partial_\phi^2\left[
\mathcal{B}(\phi)\mathcal{P}(\phi,t)\right],
 \label{eq:FPphi}
\end{equation}
with
\begin{equation}
\begin{array}{l}
 \mathcal{A}(\phi)=-\frac{\epsilon^2}{2} \phi (1-\phi^2)\\[12pt]
 \mathcal{B}(\phi)=\frac{1}{N}(1-\epsilon^2)(1-\phi^2).
\end{array}
\label{eq:FPapprox}
\end{equation}
Eq. (\ref{eq:FPphi}) is equivalent to the following Langevin equation in the Ito prescription \cite{Gardiner}
\begin{equation}
\begin{array}{lcl}
 \dot{\phi}&=&\mathcal{A}(\phi)+\sqrt{\mathcal{B}(\phi)}\eta(t)\\[12pt]
 &=&-\frac{\epsilon^2}{2} \phi (1-\phi^2)+\sqrt{\frac{1}{N}(1-\epsilon^2)(1-\phi^2)}\eta(t),
 \end{array}
 \label{eq:LangevinPhi}
\end{equation}
where $\eta$ is a $\delta$-correlated in time gaussian white noise with zero mean.  Let us emphasize that the main effect of the quenched disorder is to generate a deterministic force which stabilizes the opinion-coexistence state, $\phi=0$. In the limit $\epsilon \to 0$ we recover the purely noise-driven VM dynamics \cite{Dornic01}, while in the opposite limit $\epsilon \to 1$, the dynamics is purely deterministic and the spins align with their corresponding random fields (i.e. $\phi=0$ on average). For $0<\epsilon<1$ the stationary solution $\mathcal{P}_s(\phi;\epsilon)$ is formally given by the zero-current condition \cite{Gardiner} \begin{equation}
 J=-\mathcal{A}(\phi)\mathcal{P}(\phi,t)+\frac{1}{2} \partial_\phi [\mathcal{B}(\phi)\mathcal{P}(\phi,t)]=0
\end{equation}
from which
\begin{equation}
 \mathcal{P}_s(\phi;\epsilon)=\frac{1}{\mathcal{Z}\mathcal{B}(\phi)}e^{2\int^\phi dx \frac{\mathcal{A}(x)}{\mathcal{B}(x)}}.
\end{equation}
$\mathcal{Z}$ is supposed to be the normalization constant of $\mathcal{P}_s(\phi;\epsilon)$, but since the diffusion term $\mathcal{B}(\phi)\to 0$ when $\phi\to\pm 1$ and the exponential stays finite, the probability distribution is \emph{not normalizable} (as corresponds to the probability distribution collapsing to one of the absorbing states \cite{Munoz98}).  Therefore, as expected, for any finite value of $N$ the only steady state is an absorbing/consensus one: coexistence is always killed on the large time limit.

\begin{figure}[htp]
  \centering
 \includegraphics[width=0.7\textwidth]{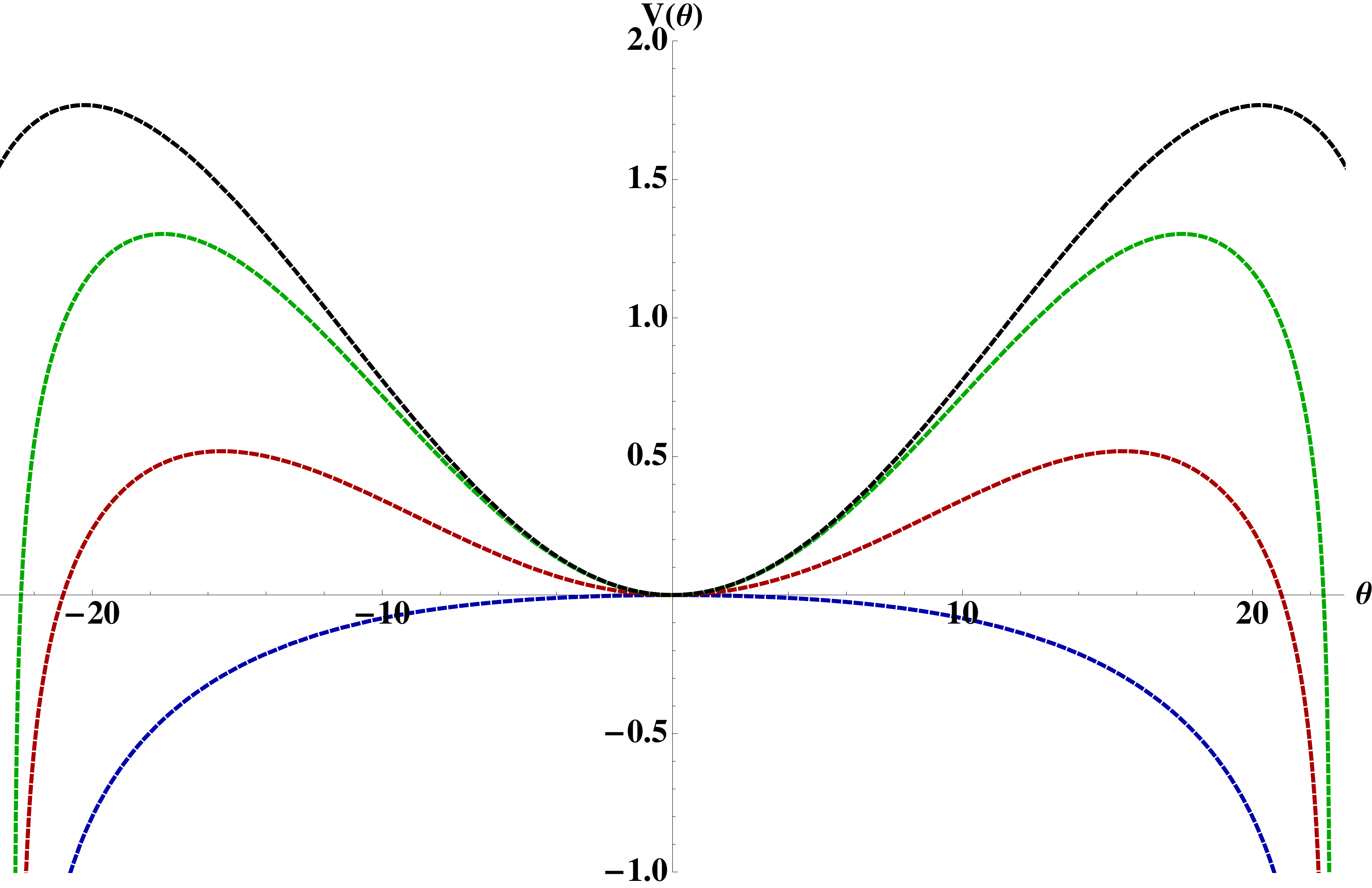}
 \caption{Potential $V(\theta)$ of Eq. (\ref{eq:potential}) corresponding to three values of $\epsilon$ at fixed $N=200$: $\epsilon=0.05$ (lower blue curve) corresponds to the absorbing phase, $\epsilon=0.15$ (red central curve) and $\epsilon=0.2$ (upper green curve) correspond to two examples of intermediate phase. The black curve corresponds to the parameters $\epsilon=0.2,N=250$. Note how the potential well becomes deeper as $\epsilon$ or $N$ increase, diverging in the $\epsilon>0,N\to \infty$ limit.}
 \label{fig:potential}
\end{figure}
The understand this problem even more transparently, we perform a change of variables
 on the Fokker-Planck equation (\ref{eq:FPphi}) such that its corresponding Langevin equation (\ref{eq:LangevinPhi}), characterized by a state-dependent (multiplicative) noise, becomes a new state-independent equation, i.e. the noise becomes additive rather than multiplicative. A suitable transformation is \cite{Baxter07,Russell11} \begin{equation}
  \theta\equiv \frac{1}{\alpha}\arcsin \phi
\end{equation}
with $\alpha=\sqrt{\frac{1}{N}(1-\epsilon^2)}$, which leads to the following Langevin equation
\begin{equation}
 \dot{\theta}=-\frac{dV(\theta)}{d\theta}+\xi,
\end{equation}
where 
\begin{equation}
 V(\theta)=\frac{1}{2} \log(\cos(\alpha  \theta ))+\frac{\epsilon ^2}{4 \alpha ^2}\sin^2(\alpha  \theta ).
 \label{eq:potential}
\end{equation}
is a potential and $\xi$ is a standard $\delta$-correlated Gaussian white noise.  Observe that the first term destabilizes the $\theta=0$ (coexistence) solution, while the second, disorder-induced, one stabilizes it (indeed, the second term in eq. (\ref{eq:potential}) can be expanded around $\theta=0$, leading to a parabolic potential around the origin; i.e. disorder creates an effective potential whose minimum corresponds to the opinion-coexistence state).  $V(\theta)$ is shown in Figure \ref{fig:potential} for some particular values of the parameters $\epsilon$ and $N$. We can see that there exists a critical value of $\epsilon$, $\epsilon_c$ (to be computed later) below which the potential effectively pushes the system toward the absorbing boundaries. Going above this critical value a local minimum appears in the configuration of zero magnetization $\phi=0$. We expect then that, for $\epsilon>\epsilon_c$ the time needed to reach the absorbing state will be exponential in the height of the potential barrier, due to the Arrhenius law \cite{Gardiner}.

As $\epsilon$ and/or $N$ increase, the basin of attraction of the minimum at $\phi=0$ becomes larger and deeper, that is, the two symmetric maxima of the potential become closer to the corresponding absorbing states and the barriers become higher, in such a way that the time needed to escape the barrier becomes much longer (actually, long enough to make it unaccessible to computer simulations). We can then identify three different regimes that we call the \emph{absorbing}, \emph{intermediate} (quasi-active) and \emph{active} phase, respectively. In the absorbing phase symmetry is broken and one of the two possible states of consensus is reached with certainty; the active phase is characterized by a coexistence of both states (it survives to fluctuations only in the infinite size limit); finally, the intermediate state is a mixture of the two previous ones and it is found for $\epsilon>\epsilon_c$ and $N<\infty$: both the consensus state and the coexistence one are locally stable, thus, the system is tri-stable, and the steady state depends on initial conditions.  

These results provide a nice illustration of how noise can effectively change the shape of the deterministic potential, allowing for noise-induced phenomena.  Still, the presence of absorbing states -- with the associated singularities in the steady state distribution -- hinders true phase transitions to occur: the only possible steady state for any finite system is an absorbing one. Instead, in the infinite size limit, noise vanishes and the coexistence state becomes truly stable.

\subsection{Finite-$N$ approximate solution: Introducing a ``mutation'' rate}
\label{sec:mutation}
In order to regularize the singularities reported above and explore the possibility of phase transitions, we introduce a small ``mutation'' term.  Mutation is defined as the process by which any randomly selected spin spontaneously inverts its state (regardless of its neighbors or external field) at some rate $\nu\gtrsim 0$.  The transition rates $W'$ become
 \begin{equation} \begin{array}{l}
 W'^b_x=(1+\epsilon)\left[(1-\nu)(\oh-x)(\oh+x-y)\right]+\frac{\nu}{2}(\oh-x)\\
 W'^d_x=(1-\epsilon)\left[(1-\nu)x(\oh-x+y)\right]+\frac{\nu}{2}x\\
 W'^b_y=(1+\epsilon)\left[(1-\nu)(\oh-y)(\oh+y-x)\right]+\frac{\nu}{2}(\oh-y)\\
 W'^d_y=(1-\epsilon)\left[(1-\nu)y(\oh-y+x)\right]+\frac{\nu}{2}y.
\end{array}
\label{eq:B-Dnuepsilon}
\end{equation}
Assuming $\nu\ll\epsilon\ll 1$ and keeping the leading orders in $\nu$ and $\epsilon$, the Fokker-Planck
equation for the global magnetization in presence of speciation becomes 
\begin{equation}
  \dot{\mathcal{P}}_\nu(\phi, t)=-\partial_\phi \left[ \mathcal{A}_\nu(\phi)\mathcal{P}_\nu(\phi,t)\right]+\frac{1}{2} \partial_\phi^2\left[
\mathcal{B}_\nu(\phi)\mathcal{P}_\nu(\phi,t)\right]
 \label{eq:FPnu}
\end{equation}
with
\begin{equation}
\begin{array}{c}
 \mathcal{A}_\nu(\phi)=-\frac{\epsilon^2}{2} \phi (1-\phi^2+2\nu)\\[12pt]
 \mathcal{B}_\nu(\phi)=[(1-\epsilon^2)(1-\phi^2)+2\nu]/N,
\end{array}
\label{eq:ABnu}
\end{equation}
respectively, and the associated stationary probability distribution function becomes
\begin{equation}
  \mathcal{P}^\nu_s(\phi;\epsilon)\propto \frac{1}{ \left(1-\epsilon ^2\right) \left(1-\phi ^2 \right)+2 \nu} \exp \left(-\frac{N}{2}\frac{\epsilon ^2 }{1-\epsilon
      ^2}\phi ^2\right),
  \label{eq:stationary}
\end{equation}
which, owing to $\nu$, does not have any singularity \cite{Nota2}. It is important to notice that if $\nu$ is small enough, namely $\nu\ll 2/(2+N)$ (see Appendix), 
it does not affect significantly the dynamics, apart from removing the absorbing boundaries. 

\begin{figure}[htp]
  \centering
 \includegraphics[width=0.7\textwidth]{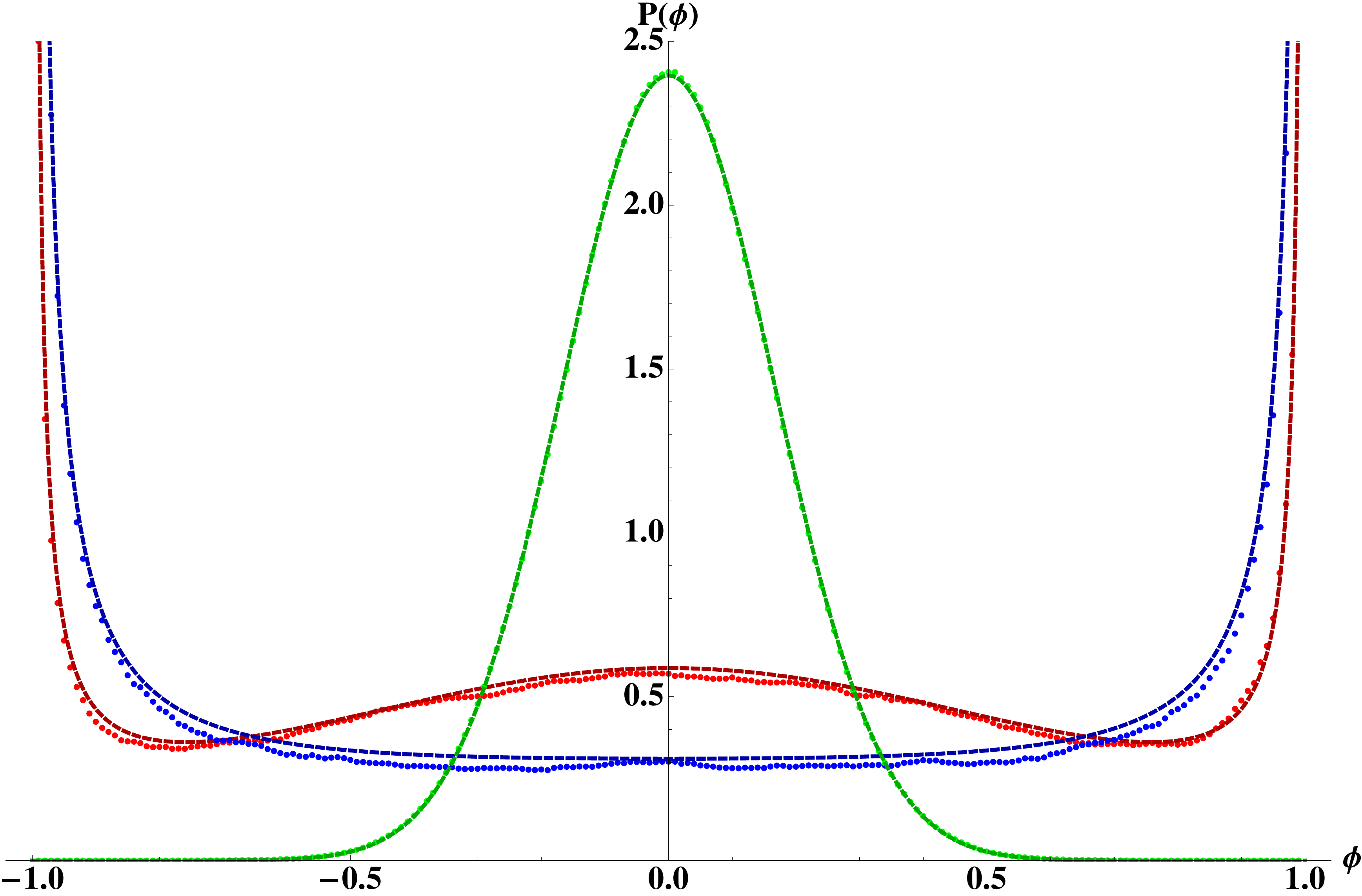}
 \caption{
Stationary probability distribution $\mathcal{P}_s(\phi; \epsilon)$ in the case with mutation, $\nu=10^{-4}$, for a finite complete network of size  $N=200$ and different values of the disorder strength,
\emph{Dashed Lines:} Curves  for, $\epsilon=0.09<\epsilon_c\simeq 0.1$
(blue curve),   $\epsilon=0.15>\epsilon_c$ (red curve), and $\epsilon=0.4$ (green curve) as computed analytically from Eq. (\ref{eq:stationary}).
\emph{Dots:} Numerical results for simulations of the model (fully connected network and Gillespie algorithm) for the same parameters as above.}
 \label{fig:stationary}
\end{figure}

We have used equation (\ref{eq:stationary}) to check our (approximate) results against numerical simulations of the complete (exact) dynamics (as obtained for a complete graph of $N$ spins by means of the Gillespie algorithm \cite{Gillespie77}). Results for different values of $\epsilon$, are reported in Figure \ref{fig:stationary} which shows a rather good agreement with the theoretical predictions.  
Finally, from equation (\ref{eq:stationary}) it is easy to compute the value of $\epsilon$ at which the second derivative of $\mathcal{P}^\nu_s$ computed in $\phi=0$ changes sign in the small $\nu$ limit: \begin{equation}
 \epsilon_c\simeq\sqrt{\frac{2}{2+N}},
\end{equation}
signaling a bifurcation from a unimodal distribution (coexistence) to a bimodal one.
However, this is not a true thermodynamic phase transition; as expected, $\epsilon_c\to 0$ when $N\to \infty$. It is interesting to notice that the bifurcation from the absorbing to the active phase is completely 'noise-driven': any arbitrarily small amount of quenched-noise --i.e. any value of $\epsilon>0$-- leads to a stable active phase.  Summing up: once absorbing states are perturbed with a non-vanishing mutation rate the only remaining stable-state in the thermodynamic limit is the active one, while for finite sizes there is a noise-induced transition.

\section{Connection to the Generalized-voter class}

Al Hammal et al. \cite{AlHammal05} introduced a phenomenological Langevin equation aimed at capturing all the possible features of systems with two-symmetric absorbing states.  Following the notation in \cite{AlHammal05} this equation reads \begin{equation}
 \dot{\phi}(x, t)=\nabla^2 {\phi}(x, t) + (a\phi-b\phi^3)(1-\phi^2)+\sqrt{1-\phi^2}\xi(x, t).
 \label{eq:alhammal}
\end{equation}
where $a$ and $b$ are constants, $\phi(x,t)$ is a field whose dynamics is frozen if $\phi^2\equiv 1$, and $\xi(x, t)$ is a Gaussian white noise of zero mean and variance $\sigma$. With the only requirement that $b < 0$ this equation reproduces the critical behavior of the GV universality class.  On the other hand, if $b>0$ then the equation resembles very much that of the Ising model (Model A), and indeed, in such a case one obtains that the GV transition is split into two different phase transitions: (i) the first one is Ising-like and corresponds to a breakdown of the up-down, $\mathbb{Z}_2$, symmetry; (ii) once the symmetry has been broken, by further changing parameters such that $\sqrt{a/b}\notin[-1, 1]$, the system eventually falls into the corresponding absorbing state (directed-percolation like phase transition). From this perspective, as pointed out in \cite{AlHammal05} the GV transition can be viewed as the merging of two different phenomena: the breaking down of a $\mathbb{Z}_2$ symmetry and the falling down into an absorbing state, both of them occurring at the same transition point.

If now we consider Eqs. (\ref{eq:FPphi}) and (\ref{eq:FPapprox}) for the description of the disordered model considered here (without mutation), we can write it in the equivalent form of the Langevin equation in the Ito prescription \cite{Gardiner} 
\begin{equation}
 \dot{\phi}(t)=-\frac{\epsilon^2}{2}\phi(1-\phi^2)+\sqrt{(1-\epsilon^2)(1-\phi^2)}\eta(t)
\label{eq:LangevinRFVM}
\end{equation}
where $\eta(t)$ is a Gaussian white noise with $\langle\eta(t)\rangle=0$ and $\langle\eta(t)\eta(t')\rangle=\frac{1}{N}\delta(t-t')$, which coincides with the 0-dimensional version of eq. (\ref{eq:alhammal}) (i.e. eq. (\ref{eq:alhammal}) without spatial dependence) once the identifications $b = 0$, $a=-\epsilon^2/2$ and $\sigma= \sqrt{\frac{1}{N}(1-\epsilon^2)}$ are made. Therefore, at least at mean-field level, the VM with quenched random field closely resembles the GV dynamics (without quenched disorder).  The main effect of random fields is to create a deterministic force which converts the state of coexistence ($\phi=0$) into a stable one. Such a state is the only possible stationary state for infinitely large systems, while for finite-sizes there is a transition very similar to that of the GV class, without quenched disorder.  Then, it is somewhat surprising that also this model is effectively well described by the same equations. 

 It is also noteworthy that $\epsilon=0$ corresponds to the critical point in the thermodynamic limit and then, as $\epsilon>0$, only the active phase exists: the absorbing phase of the AlHammal's equation (\ref{eq:alhammal}) is not accessible to the present model with quenched random fields. Thus, it is interesting to investigate what happens when quenched-disorder is introduced into a (non-linear) version of the VM, including an active and an absorbing phase, and a critical point separating the two of them. We tackle this problem in the next section.

\section{Nonlinear voter models, disorder, and spontaneous symmetry breaking}

So far we have seen that quenched disorder pushes a linear VM out of the criticality introducing an effective potential term that forces the system into an active symmetric phase.  In this section we extend the analysis to the larger class of \emph{nonlinear voter models} (NV) by mean field analysis and simulations of the model introduced in \cite{Borile12} in presence of a disordered environment. 

A general argument by Imry and Ma \cite{Imry75} predicts that, at equilibrium, quenched disorder prevents the spontaneous symmetry breaking of a discrete symmetry in $D\leq 2$ and of a continuous symmetry in $D\leq 4$ \cite{Nota1} (a rigorous proof has been given later in \cite{Aizenman89}).  However, in a remarkable recent paper Barghathi and Vojta \cite{Vojta12} established that a one-dimensional model in the GV class violates the Imry-Ma result: owing to its non-equilibrium nature it can exhibit a spontaneous symmetry breaking even in low dimensions ($D=1$ in their work).

Thus, we want to study the effect of quenched random field disorder in the remaining case, i.e. when the non-linear voter model is such that it exhibits two separate transitions occurring separately. Does this scenario survive  the presence of quenched disorder? 

We consider a particular nonlinear Voter Model introduced in \cite{Borile12} where, if $x$ is the density of spins not aligned with $\sigma_i$ in its neighborhood, the flipping probability of a randomly selected spin is given by \begin{equation}
 f_i(x)=P(\sigma_i\to-\sigma_i|x)\propto xK(x),
\end{equation}
where $K(x)$ is an arbitrary nonlinear function of $x$. We consider a cubic function of the form
\begin{equation}
 K(x)=\tilde{a}x^2+\tilde{b}(\frac{3}{4}x-x^3)
\end{equation}
that yields a stochastic equation of motion of the form of equation (\ref{eq:alhammal}) with effective parameters $a=\tilde{a}/4$ and $b=\tilde{b}/16$. If we choose $b>0,a>0$, this model has two (meta)stable fixed points at global magnetization $\phi_*=\pm\sqrt{a/b}$ \cite{AlHammal05,Castellano09a}. Next, we include disorder by adding a quenched site-dependent term $\epsilon_i=-\epsilon\tau_i\sigma_i$ that locally breaks the up down symmetry favoring or disfavoring the single spin-flip. The quenched variables $\tau_i$ are defined as in Section 2.

\subsection{Mean Field theory}

\begin{figure}[htp]
  \centering
 \subfigure{\includegraphics[width=0.8\textwidth]{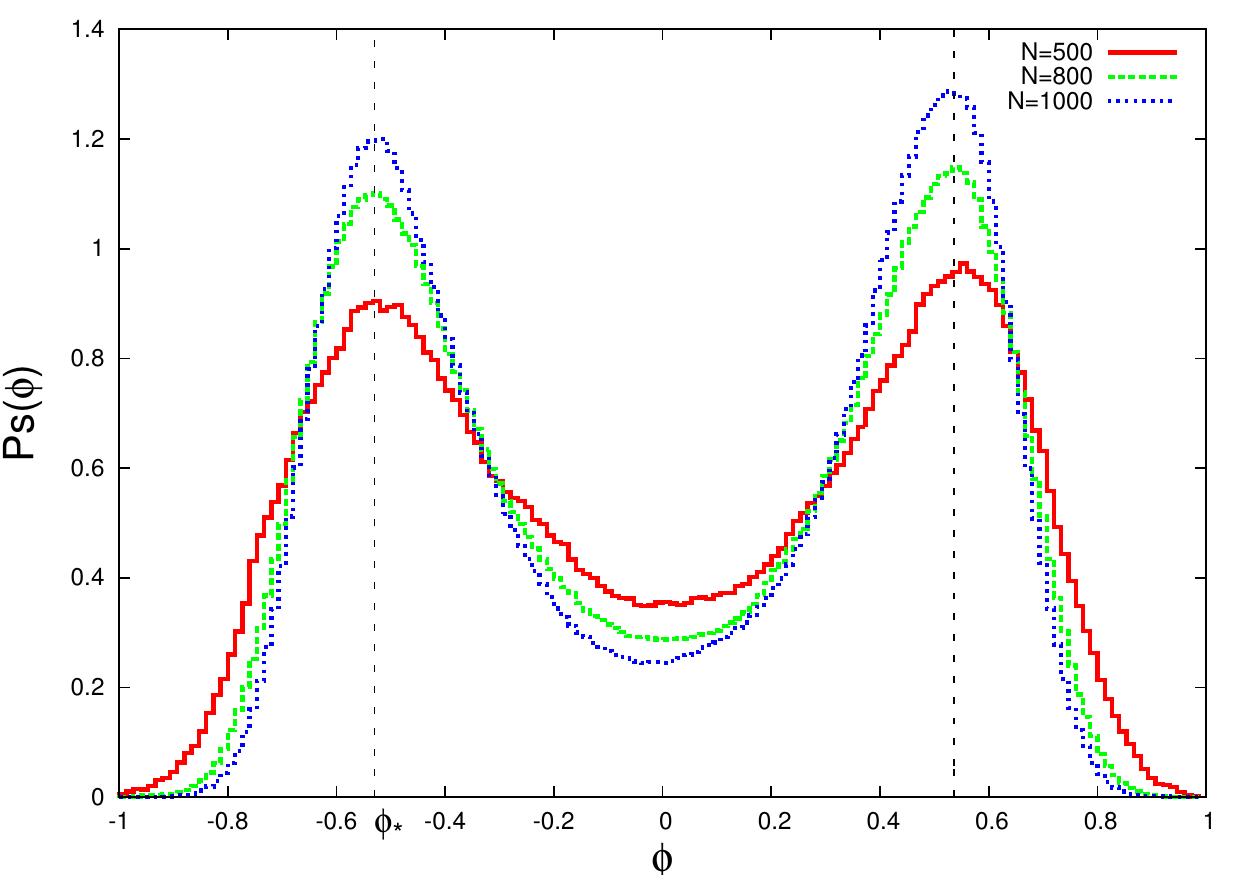}}
 \subfigure{\includegraphics[width=0.8\textwidth]{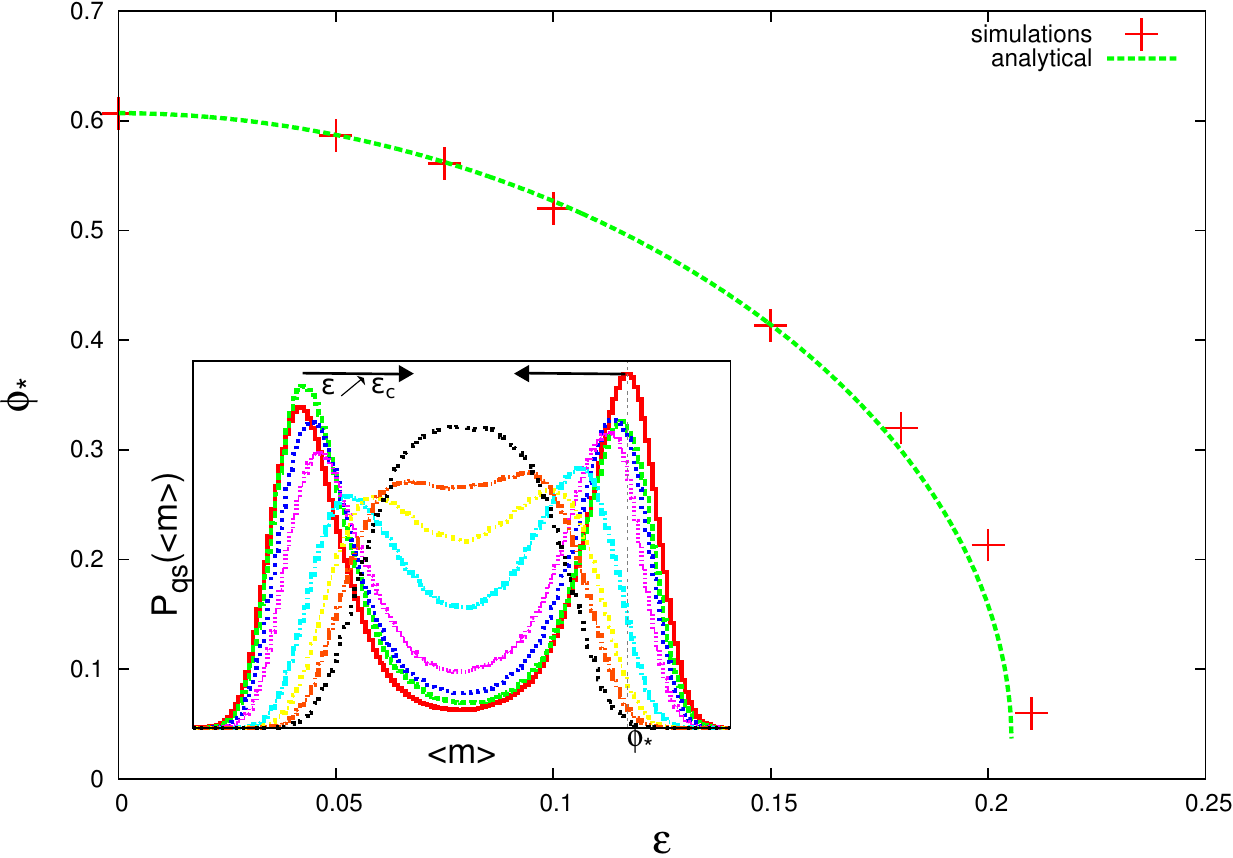}}
 \caption{Non-linear voter model running upon a fully connected network and quenched disorder of strength $\epsilon$: \emph{top:} Stationary probability distributions $P_s(\phi)$ plotted for $\epsilon=0.1<\epsilon_c^{mf}$ and various values of the system size, $N$. The position of the peak $\phi_*$ does not depend on $N$; the Ising up-down symmetry is broken. \emph{bottom:} Position of the peak $\phi_*$ for $a=0.0175/(1+\epsilon)$ and $b=0.053/(1+\epsilon)$ and varying $\epsilon$. The green dashed line represents the prediction Eq. (\ref{eq:phistar}). In the inset are shown the probability distributions for some values of $\epsilon$: as $\epsilon \rightarrow \epsilon_c$ its two peaks become closer and closer, until eventually they merge at $\epsilon_c$.  }
 \label{fig:Psmf}
\end{figure}

It can be conjectured that --in analogy with the linear case-- for this nonlinear VM mean field in a random field, with parameters of nonlinearity $a$ and $b$ and disorder strength $\epsilon$, the presence of disorder leads to an equation of the form of Eq. (\ref{eq:alhammal}), but with a new effective parameter $a'=a-\epsilon^2/2$. Namely, we expect the system to be described by \begin{equation}
 \dot{\phi}(t)=[(a-\frac{\epsilon^2}{2})\phi-b\phi^3](1-\phi^2)+\eta(\phi,t).
\label{eq:LangevinRFNV}
\end{equation}
From this equation we can compute the critical value of the external field intensity, $\epsilon^{sb}_c$, at which the disorder destroys the possibility of a spontaneous symmetry breaking, that is, $\epsilon^{sb}_c=\sqrt{2a}$. Analogously, the position of the minimum of the potential (and therefore the maximum in the stationary probability distribution of the magnetization) is expected, when $a,b>0$, at 
\begin{equation}
 \phi_*\simeq\pm\sqrt{\frac{a-\epsilon^2/2}{b}}.
 \label{eq:phistar}
\end{equation}
These results turn out to be in good agreement with computer simulations in a fully connected network; in particular, in Figure \ref{fig:Psmf} we show the dependence of the peaks position, $\phi_*$, on the intensity of the disorder and that it is independent of the size of the system.  Observe that the up-down symmetry is spontaneously broken in this mean-field like case.

\subsection{Simulations in $D=2$} 
\begin{figure}[htp]
  \centering
 \includegraphics[width=0.8\textwidth]{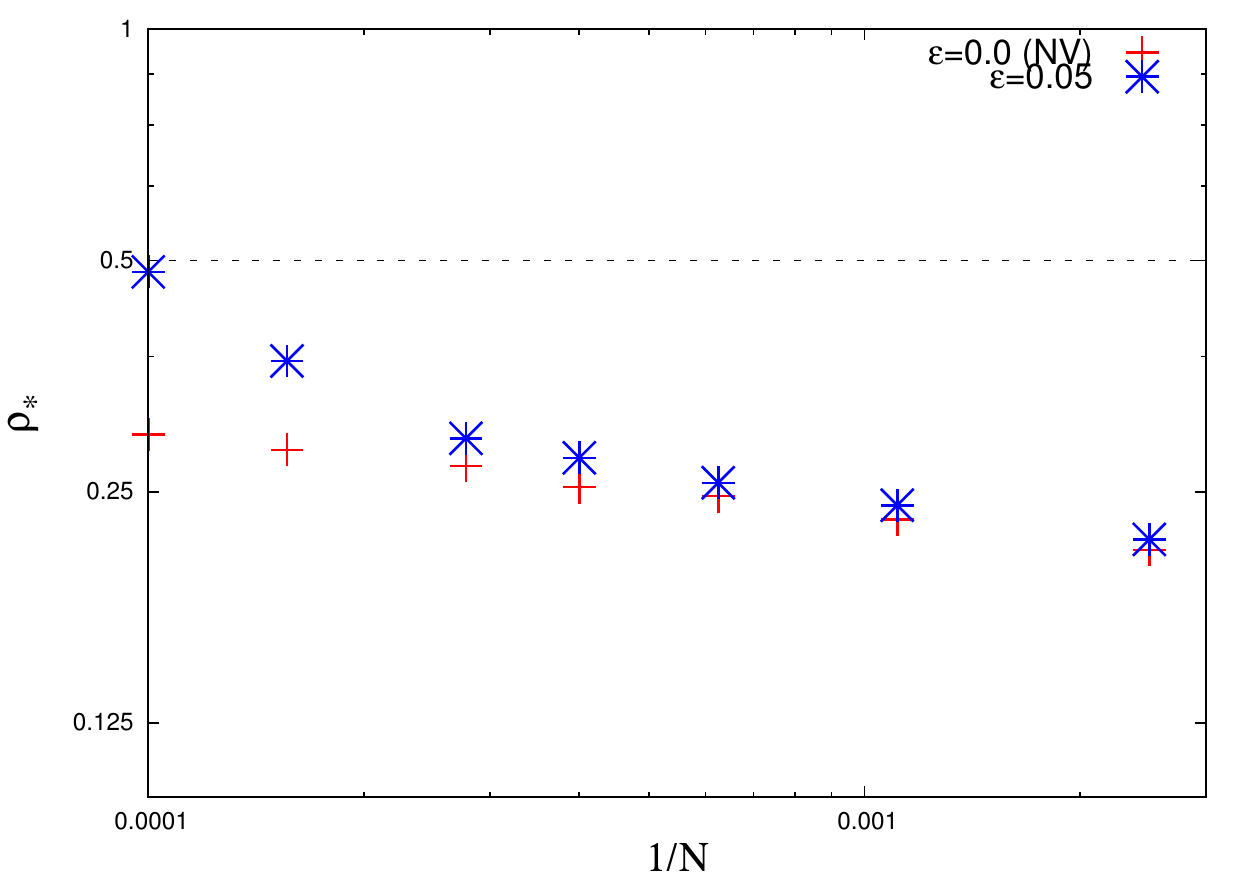}
 \caption{Two dimensional results for the non-linear voter model: position of the peaks as function of $1/N$ for the model with the same parameters as in figure \ref{fig:Psmf}, but in $D=2$. The dashed line at $\rho_*=(\phi_*+1)/2= 0.5$, to which the maxima of the distribution functions converge upon increasing system size, indicates that the symmetric active state is restored in the thermodynamic limit. }
 \label{fig:peaks}
\end{figure}

When $D\leq2$, as discussed above, the phenomenology could be radically different if the Imry-Ma argument holds.  In this case, the position of the peaks, $\rho_*=(\phi_*+1)/2$, depends on system size $N$.  Indeed, in Figure \ref{fig:peaks} we show results for the position of the peaks as a function of $1/N$ in two different cases, with and without disorder.  In the pure case ($\epsilon=0$) the curve converges to some value between $0.25$ and $0.3$, implying that symmetry breaking is preserved in the thermodynamic limit. Instead, as soon as $\epsilon \neq 0$, the peak position tends to its symmetric-state value $1/2$ as size is increased: the presence of disorder makes the symmetry breaking disappear for sufficiently large system sizes, in agreement with the argument of Imry and Ma for the equilibrium Ising symmetry breaking.  

Therefore, the non-linear voter model is a non-equilibrium one which--as predicted by the general theory of AlHammal et al.-- can exhibit two different scenarios depending on parameter values: (i) either $2$ separate phase transitions, Ising (for the up-down symmetry breaking) and DP (for the falling into the absorbing phase) or, instead, (ii) a unique GV transition (when up-down symmetry breaking and falling into the absorbing state occur simultaneously). In the first case, quenched disorder is a relevant perturbation, the Imry-Ma argument applies and no up-down symmetry breaking can occur in low dimensional systems (although it does occur at a mean-field level). Instead, in the second scenario, the GV is a genuine non-equilibrium phase transition, the Imry-Ma argument breaks down and the symmetry can be spontaneously broken even in low-dimensional cases.

\section{Conclusions}

Models with symmetric absorbing states mimicking neutral evolution --in its generalized sense-- have been proposed in different areas of science, from physics and chemistry where they can model the behavior of kinetic reactions, to biology, genetics, and ecology where models like the voter model --also called Moran process-- \cite{Kimura} or its dual representations --as the coalescent-- \cite{Kingman82} have been successfully applied to understand (at least at a null-model level) different empirical observations. In all these models and applications, nevertheless, external and environmental ``forces'' are typically ignored. A qualitative change in the trustworthiness of the mathematical approach to these problems needs a stronger consideration of the fact that the particles/individuals/agents in the system do not act in the vacuum, but live in an environment that might both modify the interactions amongst individuals and interact themselves with the constituents of the system. As a first step towards a more detailed understanding of the effect of quenched disorder in such situations, 
 in this paper we have studied how quenched random fields affect the dynamics in the voter model and in other similar non-linear models (in which the transition rates depend on the local state in a non-linear way).

 Our main findings are as follows: 

 \emph{i)} First, we investigated analytically the effects of a quenched random `preference field' in the standard voter model dynamics, finding --in agreement with previous results-- that the main effect of this type of noise is to generate a deterministic potential that stabilizes coexistence of opinions. We went further and found an effective stochastic description of the dynamics on a complete graph which is a function of the total number of spins $N$ and the disorder strength $\epsilon$, and considered versions of the model allowing for spontaneous flip (mutations) from one state to its opposite. We report on disorder-induced transitions in finite systems, which disappear when the system size becomes infinite, i.e. in the thermodynamic limit; this occurs owing to the non-trivial interplay between noise and deterministic drift.

\emph{ii)} We have found  a relation in terms of stochastic (Langevin) equations between a linear VM with disorder and the larger class of nonlinear voter models without disorder (the generalized voter Langevin equation, as proposed in \cite{AlHammal05}). 

\emph{iii)} Finally, we have scrutinized the validity of the Imry-Ma argument in a family of non-equilibrium non-linear voter models. This was motivated by the recent finding by Barghathi and Vojta \cite{Vojta12} that the Imry-MA argument can break down in non-equilibrium systems.  In particular, if the corresponding pure model has a unique GV type of phase transition, then the corresponding disordered model which quenched random fields, can still exhibit a spontaneous breaking of the up-down symmetry even in low dimensions. This is a genuinely non-equilibrium feature, prohibited in equilibrium systems.

On the other hand, as predicted by a general theory \cite{AlHammal05}, some non-equilibrium non-linear voter models, do not undergo a unique phase transition but a sequence of two of them (one Ising like and one DP like).
In this case, as we have shown here by means of computational studies and analytical arguments, 
the Imry-Ma criterion does hold and the up-down symmetry is not broken in low dimensional systems.

The underlying reason for this, is that in the case studied by Barghathi and Vojta when the discrete up-down symmetry breaks down the system simultaneously falls into one of the two absorbing states; thus fluctuations
are immediately extinguished. In the lack of stochasticity the Imry-Ma argument clearly breaks down as it rests
in evaluating the effects of eventual fluctuations into ordered states. Instead in the other case studied here, 
the broken symmetric state to-be would be a fluctuating one and the corresponding phase transition an Ising one. Therefore, at sufficiently large scales, the system behaves as an equilibrium one, the Imry-Ma argument holds and no actual symmetry breaking occurs in low dimensions.

We finally note that our description was limited for the sake of simplicity to only two species, but we do not expect qualitative changes for generalizations with more species \cite{Borile12}.

\section*{Acknowledgements}

C.B. and A.M. thank the Cariparo Foundation for financial support.
M.A.M. acknowledges partial support from  Junta de Andalucia, Proyecto de Excelencia, P09-FQM-4682 and from the spanish MICINN, as well as fruitful discussions with S. Pigolotti and M. Cencini.

\section*{Appendix. Stationary distribution for the VM with mutation}
Consider a 2-species voter model upon the complete graph consisting of $N$ nodes, without disorder, but with a mutation rate $\nu\geq 0$. Calling $\rho$ the density of one of the two species, then the global magnetization is $\phi=2\rho-1$ and in analogy with equation (\ref{eq:B-Dnuepsilon}) the birth-death rates describing the system's dynamics are
\begin{equation}
\begin{array}{lcl}
W_+&=&\ds (1-\nu)(1-\rho)\rho+\frac{\nu}{2}(1-\rho)\\[12pt]
W_-&=&\ds (1-\nu)\rho(1-\rho)+\frac{\nu}{2}\rho,
\end{array}
\end{equation}
from which it is easy to obtain a Fokker-Planck equation for the probability distribution of the global magnetization, namely
\begin{equation}
\begin{array}{rl}
\dot{P}(\phi, t)=\ds &-\partial_\phi\{ -\frac{\nu}{2}\phi P(\phi, t)\}+\\[12pt]
&+\ds\frac{1}{2N}\partial^2_\phi\{[(1-\nu)(1-\phi^2)+\nu] P(\phi, t)\}.
\end{array}
\end{equation}
Since the process is one-dimensional, it is straightforward \cite{Gardiner} to write the equation for the stationary distribution of $\phi$, 
\begin{equation}
P_\infty(\phi;N, \nu)\propto \ds \frac{1}{(1-\nu)(1-\phi^2)+\nu}e^{\nu N \int^\phi dx \frac{x}{(1-\nu)(1-x^2)+\nu}},
\end{equation}
that can be imagined as derived from a potential $V(\phi; N, \nu)$. Given a system size $N$, the distribution, initially peaked at the boundaries $\phi=\pm 1$, starts to have a peak at $\phi=0$ when
\begin{equation}
\ds \nu=\frac{2}{2+N},
\end{equation}
that can be easily verified computing the second derivative $\partial_\phi^2 P_\infty(\phi;N, \nu)|_{\phi=0}$.

Thus, for finite-size systems, when $\nu\ll 2/(2+N)$ the introduction of a mutation rate has the effect of removing the absorbing points of the process and it justifies the assumption in section \ref{sec:mutation}. It is worth noting that the Langevin equation of this process, with mutation but without disorder, is very close to equation (\ref{eq:LangevinPhi}) if $\nu\equiv \epsilon^2$. This backs the intuition that the quenched external field acts effectively as a diffuse source of particles (individuals) in the same way as a mutation term does, but with the substantial difference that while mutation eliminates the absorbing states, this does not happen in the case of the disordered system.

\section*{References}
\bibliographystyle{unsrt}
\bibliography{Bibliography}

\begin{thebibliography}{10}
\bibitem{Nota2}
This regularization is similar in spirit to what happens with the spontaneous
  magnetization of equilibrium spin systems: An external magnetic field is
  introduced in order to explictly break the symmetry, and it is turned of only
  after the thermodinamic limit of the system is taken. Here the mutation term
  is introduced in order to remove the singularity at the boundaries, and the
  active steady state is recovered in the infinite size system by applying the
  limits in the correct order, namely $\lim_{\nu\to
  0}\lim_{N\to\infty}P_s^\nu(\phi, \epsilon)=\delta(\phi)$, $\forall\
  \epsilon>0$.
\bibitem{Nota1}
Consider an Ising model in a quenched zero-mean, uniformly distributed random
  field $h$, $ H=-J\sum_{\langle i,j\rangle} \sigma_i \sigma_j+\sum_i
  h_i\sigma_i$, and consider the stability of the ferromagnetic state, all spin
  up. If we flip all spins within a region of linear size $L$, the energy cost
  for creating a domain wall will be of order $2JL^{D-1}$. On the other hand
  this might lead an energy gain of order $2\sigma L^{D/2}$ due to the quenched
  random field $h$ where $\sigma$ is the variance of the random field
  distribution. If $D>2$ at sufficiently low temperature the ferromagnetic
  order is not destroyed by the presence of the random field whereas when $D<2$
  the ferromagnetic order is unstable. For the marginal case of $d=2$
  dimensions the prediction is that the fluctuating field will have the
  dominant effect. For systems with continuous symmetries, the domain wall
  energy is of order $L^{D-2}$ so that in this case the marginal dimension is
  $4$.
\end{thebibliography}

\end{document}